\documentclass[preprint,showpacs,amsmath,amssymb]{revtex4}

\usepackage{graphicx}
\usepackage{dcolumn}
\usepackage{bm}

\begin{document}

\title{Nonrelativistic system of interacting particles in the model \\
     of the noncommutative operators of coordinates and momenta \\
     of different particles}

\author{M. V. Kuzmenko}
\affiliation{Bogolyubov Institute for Theoretical Physics of the NAS of Ukraine, \\
Metrolohichna Str., 14b, Kyiv-143, 03143, Ukraine}

\date{\today}
\begin{abstract}
It is shown that the Schr\"{o}dinger equation for a system of interacting particles 
whose Compton wavelengths are of the same order of magnitude as the system size 
is contradictory and is not strictly nonrelativistic, because it is based on the 
implicit assumption that the velocity of propagation of interactions is finite.  
In the framework of the model of the noncommutative operators of coordinates and 
momenta of different particles, the equation for a wave function which has no 
above-mentioned drawbacks is deduced. The significant differences from solutions 
of the nonrelativistic Schr\"{o}dinger equation for large values of the interaction 
constant are found, and the comparison of analogous results for hydrogenlike 
atoms with experimental data is carried out.
\end{abstract}


\maketitle

\section{Introduction}

All the known relativistic and quasirelativistic quantum descriptions 
of systems of interacting particles are constructed, as a rule, 
so that they are reduced to the well-known Schr\"{o}dinger equation 
in the nonrelativistic limit.  Therefore, there arises the question 
about the correctness of the Schr\"{o}dinger equation in the description 
of a nonrelativistic system of particles interacting with one another 
through various potentials with any admissible parameters.  The present 
work is devoted to the study of this question.

As known, the Compton wavelength $\hbar /m c$~\cite{R1} is the limit, up to 
which one can comprehensively introduce the notion of coordinate for a 
nonrelativistic particle with mass $m$.  Here, $\hbar = h/2\pi$, 
where $h$ is the Planck constant, and $c$ is the velocity of light in vacuum.  
That is, the coordinate of a nonrelativistic particle cannot be measured with 
higher accuracy than its Compton wavelength:
\begin{equation}
\Delta x > \frac{\hbar }{m c}\; .  \label{E1}
\end{equation}
Therefore, for the system of two nonrelativistic interacting particles with 
masses $m_1$ and $m_2$,  respectively, the mean distance between the particles 
cannot be measured with higher accuracy than
\begin{equation}
\Delta _{12} = \sqrt {\left( \frac{\hbar }{m_1 c} \right)^2 + \left( \frac{%
\hbar }{m_2 c} \right)^2 } = \frac{\hbar }{\mu c}\sqrt {1 - \frac{2\mu }{M}}%
\; ,  \label{E2}
\end{equation}
where $\mu$ is the reduced mass of the system, $\mu ^{ - 1} = m_1^{ - 1} + m_2^{ - 1}$,
and $M=m_1+m_2$ is the total mass of the system of two particles.  
Here, we assume that $\hbar /m_1 c$ ($\hbar /m_2 c$) is the root-mean-square 
deviation upon the measurement of the independent coordinate of the first (second) 
particle.  Then $\Delta _{12}$ in Eq.~(\ref{E2}) is the root-mean-square 
deviation upon the measurement of the distance between the particles~\cite{R2}. 
Therefore, there is no sense in the saying about the mean distance between 
the particles which is less than $\Delta _{12}$.

On the other hand, the mean distance between particles in the Schr\"{o}dinger 
theory of a hydrogenlike atom in the ground state is equal to
\begin{equation}
\left\langle \left| {\bf r}_2 - {\bf r}_1 \right| \right\rangle =
\frac{3}{2}\,\frac{\hbar}{\mu c}\,\frac{1}{\alpha Z}\; .  \label{E3}
\end{equation}
Here, $Z$ is the atomic nucleus charge and $\alpha$ is the fine structure constant.  
It is clear that we can get $\left\langle \left| {\bf r}_2 - {\bf r}_1 \right| 
\right\rangle
\ll \hbar / {\mu c}$ (for a hydrogenlike atom, $\mu /M \ll 1$)  
for sufficiently large $Z$.  In this case, we do not consider the 
question about the rightfulness of the nonrelativistic approximation.

These contradictions indicate that the Schr\"{o}dinger equation for a system 
of nonrelativistic interacting particles is not fully correct because the 
mean distance between particles can be significantly less than $\Delta _{12}$.

The probabilistic treatment of the squared modulus of a wave function 
is possible only under the assumption that the measurements of coordinates 
or momenta of different particles do not basically perturb each other even 
in the presence of any interaction forces between particles~\cite{R3}.  
This means that the operators of coordinates or momenta of two 
particles commute with each other.  
In addition, the operators of coordinates and momenta of different 
particles commute with one another in the Schr\"{o}dinger theory, which 
means the absence of mutual interferences upon the measurements of 
a coordinate of one particle and a momentum of the other.  
The last assertion is true if the time of measurement of the coordinate 
of a particle is considerably less than the time of propagation of a 
light signal across the distance equal to the system size or, 
which is the same, if the Compton wavelengths of particles are 
considerably less than the system size.  Therefore, the Schr\"{o}dinger 
equation works very well in atomic physics and solid state physics.  
However, the application of the Schr\"{o}dinger equation to atomic 
nuclei seems to be not exactly correct, because the Compton wavelength 
of a nucleon is comparable with the size of an atomic nucleus itself.  
In addition, the strict nonrelativistic formulation requires to consider 
the interaction propagation velocity to be infinite, which forces us to 
consider the operators of coordinates and momenta of different particles 
to be noncommutative with one another~\cite{R4}.  This noncommutativity leads 
to the existence of a critical value of the interaction constant for 
the Coulomb potential such that the ground state of a hydrogenlike atom 
cannot exist for its larger values.  Work~\cite{R4} is phenomenological to a 
certain extent, because the choice of the parameter inherent in the 
theory is ambiguous.  The present work develops the ideas put forth in 
the previous one~\cite{R4}. We propose to choose the parameter inherent in 
our theory in such a way that the least average distance between 
particles in the ground state in the Coulomb field be equal to $\Delta _{12}$.  
For other potentials, especially short-range ones of the Yukawa or Hulth\'{e}n type, 
the use of this parameter leads to average distances being at least $\Delta _{12}$.

\section{	Two-body problem in the framework of the model of the noncommutative 
operators of coordinates and momenta of different particles.}

As known, the classical equations of motion for a particle of mass $m$   
in the external field $V\left( {\bf r} \right)$ are derived from 
the Hamilton function
\begin{equation}
H\left( {\bf r},{\bf p} \right) = \frac{{\bf p}^2}{2 m} + V\left( {\bf r}
\right) \label{E4}
\end{equation}
which depends on the coordinates of a particle ${\bf r}$ and the corresponding 
momentum ${\bf p}$. The total energy of the system is
\begin{equation}
E = H\left( {\bf r},{\bf p} \right)\; .  \label{E5}
\end{equation}
With this classical system, we associate a quantum system whose 
dynamical state is represented by a wave function $\Psi \left( {\bf r},t \right)$
defined in the configurational space.  The wave equation is constructed 
by the formal substitution of the quantities $E$, ${\bf r}$, and ${\bf p}$ on 
both sides of relation~(\ref{E5}) by the relevant operators~\cite{R5}
\begin{eqnarray}
&& E \to \hat E  =  i\hbar \frac{\partial }{\partial t}\; ,  \label{E6}\\
&& {\bf r} \to {\bf \hat r}  =  {\bf r}\; ,  \label{E7}\\
&& {\bf p} \to {\bf \hat p}  =  - i\hbar \bm{\nabla} \; .  \label{E8}
\end{eqnarray}
It is implied that the results of action of both sides of 
equality~(\ref{E5}) considered as operators on $\Psi \left( {\bf r},t \right)$ 
are identical. In view of this fact, we get the nonrelativistic Schr\"{o}dinger 
equation for a particle in the external field $V\left( {\bf r} \right)$:
\begin{equation}
i\hbar \frac{\partial }{\partial t}\Psi \left( {\bf r},t \right) = \left[ -
\frac{\hbar ^2 }{2m}\Delta + V\left( {\bf r} \right) \right]\Psi \left( {\bf %
r},t \right)\; .  \label{E9}
\end{equation}
We emphasize that the operators ${\bf \hat r}$ and ${\bf \hat p}$ 
in Eqs.~(\ref{E7}) and (\ref{E8}) are 
written in the configurational space and ${\bf r}$ is the vector of 
a position of the particle in a rectangular coordinate system.

The operators of coordinate and momentum are noncommutative,
\begin{equation}
\left[ {\hat x \, , \, \hat p_x } \right] = i\hbar \; , \quad \left[ {\hat y%
\, ,\,\hat p_y } \right] = i\hbar\; ,\quad \left[ {\hat z\, ,\, \hat p_z } %
\right] = i\hbar\; ,  \label{E10}
\end{equation}
which yields the Heisenberg uncertainty relations
\begin{equation}
\Delta x\Delta p_x \ge \hbar/2\; , \; \Delta y\Delta p_y \ge \hbar /2\; ,
\; \Delta z\Delta p_z \ge \hbar /2\; ,  \label{E11}
\end{equation}
where the quantities $\Delta x$, $\Delta p_x$, $\Delta y$, $\Delta p_y$, $\Delta z$,
and $\Delta p_z$ are directly 
connected with relevant measurements and are the root-mean-square 
deviations from the mean value.  For example, for the coordinate $x$,  we have 
\begin{equation}
\Delta x = \sqrt {\langle \hat x^2 \rangle - \langle \hat x\rangle ^2 }\; ,
\label{E12}
\end{equation}
where $\langle \hat A\rangle$ is the mean value of the operator $\hat A$ 
on the wave function $\Psi \left( {\bf r},t \right)$.
	
Relations~(\ref{E11}) state that a particle cannot be in the states where 
its coordinate and momentum take simultaneously quite definite, exact values.  
In addition, quantum theory states that the unpredictable and uncontrolled 
disturbance undergone by the physical system in the process of measurement 
is always finite and such that the Heisenberg uncertainty relations~(\ref{E11})
are satisfied~\cite{R5}. Hence, none experiment can realize the simultaneous 
exact measurement of the coordinate and momentum of a particle.  
For example, the measurement of the coordinate $x$ with accuracy $\Delta x$  
in the well-known experiment with the use of a Heisenberg microscope 
is accompanied by the uncontrolled momentum transfer to the particle 
which is characterized by the uncertainty
\begin{equation}
\Delta p_x \approx\frac{\hbar }{2\Delta x}\; .  \label{E13}
\end{equation}
In this case, the limits of exactness in the determination of 
a position are always set by optical resolving power conditioned by 
the effects of diffraction according to classical wave optics.

If the system size is such that the characteristic time of flight
with the velocity of light across 
the system exceeds considerably the duration of the process of 
measurement of a position $\Delta t$, then we may say that the process of 
measurement of the coordinate of a particle with accuracy $\Delta x$  
is accompanied by a blow against the particle with the force
\begin{equation}
F_x \approx\frac{\Delta p_x }{\Delta t}\approx\frac{\hbar c}{2(\Delta x)^2 }\; .
\label{E14}
\end{equation}
Here, we assume that the momentum transferred to the particle under 
measurement of its coordinate is of the same order as the root-mean-square 
deviation $\Delta p_x$.

In the measurement of the momentum of a particle with accuracy $\Delta p_x$,  
it undergoes a blow with the force
\begin{equation}
F_x \approx\frac{2c}{\hbar }(\Delta p_x )^2\; .  \label{E15}
\end{equation}

Analogously to Eq.~(\ref{E9}), one can deduce the Schr\"{o}dinger  
nonrelativistic equation for a system of two interacting 
particles whose Hamilton function is
\begin{equation}
H = \frac{{\bf p}_1^2 }{2m_1} + \frac{{\bf p}_2^2 }{2m_2} + V\left(\left|
{\bf r}_2 - {\bf r}_1 \right| \right)\; .  \label{E16}
\end{equation}
Here, ${\bf r}_1$ and ${\bf r}_2$ are the Cartesian coordinates of a position of 
two particles with masses $m_1$ and $m_2$, ${\bf p}_1$ and ${\bf p}_2$ are their 
relevant momenta, and the potential energy depends only on the 
distance between the particles.

With this classical system, we associate a quantum system 
whose dynamical state is represented by a wave function 
$\Psi ({\bf r}_1 ,{\bf r}_2 ,t)$  
defined in the configurational space. The wave equation 
is derived by means of the formal substitution of the 
quantities $E$, ${\bf r}_1$, ${\bf r}_2$, ${\bf p}_1$, and ${\bf p}_2$ 
on both sides of the relation 
analogous to Eq.~(\ref{E5}) by the corresponding operators:
\begin{eqnarray}
&&E \to \hat E = i\hbar \frac{\partial }{\partial t}\; ,  \label{E17}\\
&&{\bf r}_1 \to {\bf \hat r}_1 = {\bf r}_1\; ,  \label{E18}\\
&&{\bf r}_2 \to {\bf \hat r}_2 = {\bf r}_2\; ,  \label{E19}\\
&&{\bf p}_1 \to {\bf \hat p}_1 = - i\hbar \bm{\nabla}_1\; ,  \label{E20}\\
&&{\bf p}_2 \to {\bf \hat p}_2 = - i\hbar \bm{\nabla}_2\; .  \label{E21}
\end{eqnarray}
Then the well-known Schr\"{o}dinger nonrelativistic equation for 
a system of two interacting particles reads
\begin{equation}
i\hbar \frac{\partial }{\partial t}\Psi = \left[ - \frac{\hbar ^2}{2m_1 }\Delta _
1 - \frac{\hbar ^2}{2m_2 }\Delta _2 + V\left( \left| {\bf r}_2 - {\bf r}_1 \right|
\right) \right]\Psi\; . \label{E22}
\end{equation}
The operators ${\bf \hat r}_1$, ${\bf \hat r}_2$, ${\bf \hat p}_1$,
and ${\bf \hat p}_2$ satisfy the following commutation relations:
\begin{equation}
\left[ {\hat x_k,\hat p_{kx} } \right] = i\hbar,\,\left[ {\hat y%
_k,\hat p_{ky} } \right] = i\hbar,\,\left[ {\hat z_k,\hat p%
_{kz} } \right] = i\hbar, \; k = 1,2.  \label{E23}
\end{equation}
All the rest possible commutation relations are zero including such as
\begin{equation}
\left[ {\hat x_k,\hat p_{lx} } \right] = 0,\,\left[ {\hat y_k,
\hat p_{ly} } \right] = 0,\,\left[ {\hat z_k,\hat p_{lz} } \right]
= 0,\;(k \ne l).  \label{E24}
\end{equation}
Equalities~(\ref{E24}) are based on the assumption that the measurements 
of coordinates and momenta of different particles do not basically 
disturb one another even in the presence of any interaction forces 
between particles~\cite{R3}.  That is, it is assumed that a change of the 
force action of one particle on the other one induced by the 
measurement of the coordinate of the former is propagated with finite velocity.

Thus, in the derivation of the Schr\"{o}dinger nonrelativistic equation 
for a system of two particles, one uses, on the one hand, 
a Hamilton nonrelativistic classical function and, on the 
other hand, the implicit assumption that the interaction 
propagation velocity is finite. 

Within the fully nonrelativistic quantum theory, we must consider 
the interaction propagation velocity to be infinite, which forces 
us to refuse the fulfillment of the commutation relations~(\ref{E24}).  
From this viewpoint, we will consider that, under the measurement 
of the coordinate of the first particle, the uncontrolled momentum 
transfer to not only this particle, but to the whole system, occurs 
since the particles are bound with each other by the interaction 
potential whose propagation velocity is infinite. Therefore, it 
is natural to require that the commutator of the coordinate 
operator of the first particle and the operator of the total 
momentum of the system be equal to $i\hbar$:
\begin{equation}
\left[ {\hat x_1\, ,\,\hat P_x } \right] = i\hbar\, ,\, \left[ {\hat y_1\, ,
\,\hat P_y } \right] = i\hbar\, ,\, \left[ {\hat z_1 \, ,\,\hat P_z } \right]
= i\hbar\; .  \label{E25}
\end{equation}
Here, ${\bf \hat P} = {\bf \hat p}_1 + {\bf \hat p}_2$ 
is the operator of the total momentum of the system.  
The same should be true for the second particle:
\begin{equation}
\left[ {\hat x_2\, ,\,\hat P_x } \right] = i\hbar\, ,\, \left[ {\hat y_2\, ,
\,\hat P_y } \right] = i\hbar\, ,\, \left[ {\hat z_2\, ,\,\hat P_z } \right] =
i\hbar\; .  \label{E26}
\end{equation}
We note that relations~(\ref{E25}) and (\ref{E26}) are satisfied also 
for the Schr\"{o}dinger 
nonrelativistic equation, and just they allow one to construct the operator 
of coordinates of the center of masses of the system whose commutator with 
the operator of the total momentum of the system is equal to $i\hbar$.  
On the contrary, the fulfillment of relations~(\ref{E23}) is not obligatory for 
a system of interacting particles, and we intend to refuse it.
	
Of course, under the measurement of the coordinate of some particle 
with accuracy $\Delta x$, the system undergoes a blow with the 
force $\approx{\hbar c}/{2(\Delta x)^2 }$.  
For example, the measurement of the coordinate of a nonrelativistic 
electron with the maximally possible accuracy of order of the Compton 
wavelength $\hbar /m_{e}c=3.86\times 10^{-11} \mathop{\rm cm}$ 
is accompanied by a blow with the force 
$F_{e}\approx 6.62\times 10^{9} \mathop{\rm MeV}/ \mathop{\rm cm}$.  
For a proton, the Compton wavelength is about 
$2.10\times 10^{-14} \mathop{\rm cm}$,  and the blow force equals  
$F_{p}\approx 2.24\times 10^{16}\mathop{\rm MeV}/\mathop{\rm cm}$.
The mean interaction force between the particles in a hydrogen atom 
in the ground state 
$F_{H}\approx 1.03\times 10^{4}\mathop{\rm MeV}/ \mathop{\rm cm}$,
and $F_{D}\approx 4.50\times 10^{14} \mathop{\rm MeV} /\mathop{\rm cm}$ 
for a bound state of a deuterium nucleus.  
Therefore, we can neglect the interaction force 
between the particles 
in a hydrogen atom under the measurement of their coordinates 
because $F_H /F_e \approx 1.56 \times 10^{ - 6}$ and consider 
the operators of coordinates and momenta of different particles to be commutative. 
But the situation is different in atomic nuclei, because the ratio of the 
interparticle interaction force to the blow force is 
$F_D /F_p \approx 0.02$.
	
In the general case, we take
\begin{equation}
\left[ {\hat x_1\, ,\,\hat p_{2x} } \right] = i\hbar \hat
\varepsilon _{12}\; , \label{E27}
\end{equation}
where $\hat \varepsilon _{12}$ is some dimensionless Hermitian operator.  
Then Eq.~(\ref{E25}) yields that
\begin{equation}
\left[ {\hat x_1\, ,\,\hat p_{1x} } \right] = i\hbar (1 - \hat
\varepsilon _{12} )\; .  \label{E28}
\end{equation}
Analogously, if
\begin{equation}
\left[ {\hat x_2\, ,\,\hat p_{1x} } \right] = i\hbar \hat
\varepsilon _{21}\; , \label{E29}
\end{equation}
then
\begin{equation}
\left[ {\hat x_2\, ,\,\hat p_{2x} } \right] = i\hbar (1 - \hat
\varepsilon _{21} )\; .  \label{E30}
\end{equation}

The dimensionless Hermitian operators $\hat \varepsilon _{12}$ and 
$\hat \varepsilon _{21}$ depend in the general case on the
interparticle interaction force ${\bf F}_{12}$ and on masses $m_1$ and $m_2$.  
The operators $\hat \varepsilon _{12}$ and $\hat \varepsilon _{21}$ 
cannot depend on the direction of the 
vector ${\bf F}_{12}$, because the commutation relations for the $x$, $y$, and $z$  
components must be identical analogously to Eqs.~(\ref{E27})-(\ref{E30}), because 
the system has no distinguished directions and the independent 
variables are fully equivalent in a rectangular coordinate system.  
Therefore, the operators $\hat \varepsilon _{12}$ and $\hat \varepsilon _{21}$ 
are functions of the force modulus, i.e. of $\left| {\bf F}_{12}  \right|$:
\begin{equation}
\hat \varepsilon _{12} \equiv \hat \varepsilon _{12} \left(m_1,m_2,\left| {\bf
F}_{12} \right|\right),\, \hat \varepsilon _{21} \equiv \hat
\varepsilon _{21} \left(m_1,m_2,\left| {\bf F}_{12} \right|\right).
\label{E31}
\end{equation}

For the operators $\hat x_1$ and $\hat p_{2x}$ which do not commute with each other, 
the uncertainty relation looks~\cite{R6} as
\begin{equation}
\Delta x_1 \Delta p_{2x} \ge \frac{\hbar }{2} \left| {\langle \hat
\varepsilon _{12} \rangle } \right|\; ,  \label{E32}
\end{equation}
where $\langle \hat \varepsilon _{12} \rangle$ is the 
quantum-mechanical mean in the state $\Psi ({\bf r}_1,{\bf r}_2 ,t)$.  
In the general case, the right-hand side of the uncertainty relation~(\ref{E32}) 
takes different values for every quantum state, which hampers 
significantly the derivation of a wave equation.  
The problem can be considerably simplified if the operator 
$\hat \varepsilon _{12}$ in Eq.~(\ref{E27}) is substituted by 
the modulus of its quantum-mechanical mean $\left| {\left\langle 
{\hat \varepsilon _{12} } \right\rangle }\right|$  
in the ground state of the system.  In this case, 
the right-hand side of relation~(\ref{E32}) takes the value which is 
maximum of all the possible ones, because the mean interparticle 
interaction force in the ground state is maximum and therefore 
the momentum transferred to the second particle under the measurement 
of the coordinate of the first one is maximum.  It is worth noting that, 
in such a statement, the uncertainty relation~(\ref{E32}) is not changed in 
the ground state of the system.  A similar simplification can be made 
also for the operator $\hat \varepsilon _{21}$, which allows us to eventually deduce 
a nonrelativistic wave equation for a system of two particles.

The commutation relations for the operators of coordinates or momenta 
of different particles remain the same as in the Schr\"{o}dinger theory,
\begin{equation}
\left[ {\hat x_1\, ,\,\hat x_2 } \right] = 0\; ,\quad \left[ {\hat p_{1x}
\, ,\,\hat p_{2x} }\right] = 0\; ,  \label{E33}
\end{equation}
which allows one to use these operators as independent variables.

Below, we write the commutation relations for all the operators 
of coordinates and momenta of the two-body problem:
\begin{eqnarray}
&&\left[ {\hat x_1\, ,\,\hat p_{1x} } \right] = i\hbar (1 -
\varepsilon _{12} )\; , \label{E34}\\
&&\left[ {\hat x_2\, ,\,\hat p_{2x} } \right] = i\hbar (1 -
\varepsilon _{21} )\; , \label{E35}\\
&&\left[ {\hat x_1\, ,\,\hat p_{2x} } \right] = i\hbar \varepsilon
_{12}\; , \label{E36}\\
&&\left[ {\hat x_2\, ,\,\hat p_{1x} } \right] = i\hbar \varepsilon
_{21}\; , \label{E37}\\
&&\left[ {\hat x_1\, ,\,\hat x_2 } \right] = 0\; ,  \label{E38}\\
&&\left[ {\hat p_{1x}\, ,\,\hat p_{2x} } \right] = 0\; .  \label{E39}
\end{eqnarray}
Analogous relations hold for the $y$ and $z$ components.  
We recall that $\varepsilon _{12}$ and $\varepsilon _{21}$ are 
the moduli of the quantum-mechanical means 
of the operators $\hat \varepsilon _{12}$ and $\hat \varepsilon _{21}$ 
in the ground state $\Psi _0 ({\bf r}_1,{\bf r}_2,t)$ of the system:
\begin{eqnarray}
&&\varepsilon _{12} = \left|\frac{\left\langle \Psi _0 \right|\hat
\varepsilon _{12} (m_1 ,m_2 ,\left| {\bf F}_{12} \right|)\left|
\Psi _0 \right\rangle }
{\left\langle \Psi _0\left | \right. \Psi _0 \right\rangle }\right|\;
,\label{E40}\\
&&\varepsilon _{21} = \left|\frac{\left\langle \Psi _0 \right|
\hat \varepsilon _{21} (m_1 ,m_2 ,\left| {\bf F}_{12}
\right|)\left| \Psi _0  \right\rangle} {\left\langle \Psi _0 \left
| \right. \Psi _0 \right\rangle }\right|\; . \label{E41}
\end{eqnarray}

We estimate now the quantities $\varepsilon _{12}$ and $\varepsilon _{21}$.  
We assume that the momentum transferred to a particle is of order of the 
root-mean-square deviation $\Delta p$. Then Eq.~(\ref{E34}) and Eq.~(\ref{E36}) yield
\begin{equation}
\Delta x_1 \Delta p_{1x} \approx\frac{\hbar }{2}(1 - \varepsilon
_{12})\; , \label{E42}
\end{equation}
\begin{equation}
\Delta x_1 \Delta p_{2x} \approx\frac{\hbar }{2} \varepsilon _{12}\;
. \label{E43}
\end{equation}
Whence we get
\begin{equation}
\varepsilon _{12} = \frac{\Delta p_{2x}}{\Delta p_{1x}
}\left(1 + \frac{\Delta p_{2x}}{\Delta p_{1x}} \right)^{
- 1}\; .  \label{E44}
\end{equation}
Here, $\Delta p_{2x}$ is the momentum transferred to the second particle under 
the measurement of the coordinate of the first one, and $\Delta p_{1x}$  
is the momentum transferred to the first particle under 
the measurement of its coordinate.  We assume further that 
the coordinate of the first particle is measured with 
the highest possible accuracy, i.e., $\Delta x_1 = \hbar /m_1 c$. In this case, 
the momentum transferred to the second particle can be estimated as
\begin{equation}
\Delta p_{2x} = \left\langle \left| {\bf F}_{12} \right|
\right\rangle \Delta t = \left\langle \left| {\bf F}_{12} \right|
\right\rangle \frac{\Delta x_1 }{c} = \left\langle \left| {\bf
F}_{12} \right| \right\rangle \frac{\hbar }{m_1 c^2 }\; .
\label{E45}
\end{equation}
Here, $\left\langle \left| {\bf F}_{12} \right| \right\rangle$ is 
the mean value of the force in a given quantum state, 
$\Delta t$ is the duration of measurement of the coordinate of the first particle.  
We assume that the momentum transferred to the second particle, Eq.~(\ref{E45}),
is small. Then the momentum which will be transferred to the first 
particle can be estimated as
\begin{equation}
\Delta p_{1x} = \frac{\hbar }{2\Delta x_1} = \frac{m_1 c}{2}
\label{E46}
\end{equation}
and $\Delta p_{2x}/{\Delta p_{1x} }$ can be written as
\begin{equation}
\frac{\Delta p_{2x} }{\Delta p_{1x}} = \xi \frac{m_2^2 }{M^2 }\; ,
\label{E47}
\end{equation}
where $\xi =2\hbar\left\langle \left| {\bf F}_{12} \right| \right\rangle
/{\mu ^2 c^3 }$. In what follows, we assume that the functional 
dependence on the mean interaction force $\left\langle 
\left| {\bf F}_{12} \right| \right\rangle$ in Eq.~(\ref{E47}) 
is preserved, and a more exact dependence on the masses of 
interacting particles is taken into account by the introduction 
of a constant $\Omega$ which will be defined by the requirement that 
the least mean distance between particles in the ground state of 
a hydrogenlike atom be equal to $\Delta _{12}$. 
By running ahead, we note that $\Omega$  
depends only on the ratio $\mu/M$.  Finally, we get the following 
expression for the noncommutativity parameter $\varepsilon _{12}$:
\begin{equation}
\varepsilon _{12} = \Omega \xi \frac{m_2^2}{M^2 }\left( 1 + \Omega
\xi \frac{m_2^2 }{M^2 } \right)^{ - 1}\; .  \label{E48}
\end{equation}
The noncommutativity parameter $\varepsilon _{21}$ 
can be derived analogously as
\begin{equation}
\varepsilon _{21} = \Omega \xi \frac{m_1^2 }{M^2 }\left( 1 +
\Omega \xi \frac{m_1^2 }{M^2 } \right)^{ - 1}\; .  \label{E49}
\end{equation}

Now we can construct one of the possible representations for 
the operators of coordinates and momenta of a system of two particles:
\begin{eqnarray}
&&{\bf \hat r}_1 = {\bf r}_1\; ,  \label{E50}\\
&&{\bf \hat r}_2 = {\bf r}_2\; ,  \label{E51}\\
&&{\bf \hat p}_1 = - i\hbar (1 - \varepsilon _{12} )\bm{\nabla}_1 -
i\hbar \varepsilon _{21} \bm{\nabla}_2\; ,  \label{E52}\\
&&{\bf \hat p}_2 = - i\hbar \varepsilon _{12} \bm{\nabla}_1 -
i\hbar (1 - \varepsilon _{21} )\bm{\nabla} _2\; .  \label{E53}
\end{eqnarray}
It is easy to verify that operators (\ref{E50})-(\ref{E53}) satisfy 
the commutation relations~(\ref{E34})-(\ref{E39}).

The operator of the total momentum of the system
\begin{equation}
{\bf \hat P} = {\bf \hat p}_1 + {\bf \hat p}_2 = - i\hbar
\bm{\nabla}_1 - i\hbar \bm{\nabla}_2\; .  \label{E54}
\end{equation}
By substituting the quantities in the Hamilton function~(\ref{E16}) 
by operators~(\ref{E50})-(\ref{E53}), we get the nonrelativistic wave 
equation for a system of two particles as
\begin{equation}
i\hbar \frac{\partial }{\partial t}\Psi \left( {\bf r}_1 ,{\bf r}
_2 ,t \right) = H\Psi \left( {\bf r}_1,{\bf r}_2,t \right)\; ,
\label{E55}
\end{equation}
with the Hamiltonian of the system
\begin{equation}
H = -{A}_1 \Delta _1 - {A}_2 \Delta _2 - {A}_{12} (\bm{\nabla}_1 \cdot
\bm{\nabla}_2 ) + V\left( \left| {\bf r}_1 - {\bf r}_2  \right|
\right)\; ,  \label{E56}
\end{equation}
where 
$ A_1 = {\hbar ^2}(1 - \varepsilon _{12} )^2/{2 m_1} + 
{\hbar ^2}\varepsilon _{12}^2/{2 m_2}$,
$ A_2 = {\hbar ^2}(1 - \varepsilon _{21} )^2/{2 m_2}  + 
{\hbar ^2}\varepsilon_{21}^2/{2 m_1}$,
and
$ A_{12} ={\hbar ^2}\varepsilon _{21} (1 - \varepsilon _{12} )/m_1 +
{\hbar ^2}\varepsilon _{12} (1 - \varepsilon _{21} )/m_2$.

Consider the Hamiltonian of an isolated system which does 
not depend on time, and therefore the energy of the system 
is the integral of motion.  By using the substitution
\begin{equation}
\Psi = \psi \exp\left( - i\frac{E_0 t}{\hbar } \right)\; ,
\label{E57}
\end{equation}
where $\psi$ depends on coordinates of the configurational space 
but is independent of time, we get the equation for the 
stationary states of a system of two particles $H\psi = E_0 \psi$. Here,  
$E_0$ is the total energy of the two-particle system.  
If we use the substitution of variables
\begin{equation}
{\bf r} = {\bf r}_1 - {\bf r}_2\; ,  \label{E58}
\end{equation}
\begin{equation}
{\bf R} = \frac{m_1}{M}{{\bf r}_1} + \frac{m_2}{M} {{\bf r}_2} +
\frac{m_1 \varepsilon _{12} - m_2 \varepsilon _{21} }{M \beta}\left( {\bf r}_1 - {\bf
r}_2  \right)\; , \label{E59}
\end{equation}
the equation for the wave function admits the separation of variables 
after the substitution $\psi ({\bf r},{\bf R}) = \Phi ({\bf R})\phi 
({\bf r})$. In this case, the Hamiltonian of the system takes the following form:
\begin{equation}
H = - \frac{\hbar ^2}{2M}\Delta _{\bf R}  - \frac{\hbar ^2 \beta^2}{2\mu }\Delta _{\bf r}  + V\left( \left| {\bf r} \right|
\right)\; .  \label{E60}
\end{equation}
Here, $\beta = 1 - \varepsilon _{12} - \varepsilon _{21}$.
In this case, we get the quantum-mechanical description of 
two noninteracting fictitious particles, the first of which 
represents a free motion of a particle with mass equal to the 
sum of masses of the particles and with momentum equal to the 
total momentum of the system  (${\bf \hat P} = {\bf \hat p}_1 + 
{\bf \hat p}_2 = - i\hbar \bm{\nabla}
_1 - i\hbar \bm{\nabla}_2 = - i\hbar \bm{\nabla}_{\bf R}$). 
The position of this particle 
is set by the vector ${\bf R}$ which does not define the coordinate 
of the center of masses in the general case, whereas this is 
true for the Schr\"{o}dinger equation, but only in the case of 
identical particles.  The second fictitious particle with mass  
$m = \mu \beta^{ - 2}$ moves in the field $V\left( \left| {\bf r} \right| 
\right)$ and represents the relative motion of 
two particles with energy $E$.

Since no external fields act on the system, its Hamiltonian 
must be invariant with respect to both a parallel translation 
of the coordinate system in space and a rotation of the 
coordinate axes.  In addition, the equations of motion do 
not vary under a uniform and rectilinear motion of the 
system (the Galilei invariance).
	
The operator of the total momentum of the system ${\bf \hat P} = {\bf 
\hat p}_1 + {\bf \hat p}_2 = - i\hbar \bm{\nabla}
_1 - i\hbar \bm{\nabla}_2$ is 
connected with the operator of the infinitesimal translation 
transforming the function $\Psi ({\bf r}_1 ,{\bf r}_2 )$ into 
$\Psi ({\bf r}_1 + \delta {\bf r},{\bf r}_2 + \delta {\bf r})$,
\begin{equation}
1 + \delta {\bf r}\cdot\sum\limits_{i = 1}^2 \bm{\nabla}_i = 1 +
\frac{i}{\hbar } \delta {\bf r\cdot \hat P}\; ,  \label{E61}
\end{equation}
and commutes with Hamiltonian (\ref{E60}),
\begin{equation}
\left[H\, ,\,{\bf \hat P} \right] = 0\; ,  \label{E62}
\end{equation}
where $\delta {\bf r}$ is the vector of an infinitesimal parallel translation of 
all the radius-vectors of the particles by the same value, ${\bf r}_i \to 
{\bf r}_i + \delta {\bf r}$,  
and the operator of the total momentum looks as ${\bf \hat P} 
= - i\hbar \bm{\nabla}_{\bf R}$ in variables~(\ref{E58}) and (\ref{E59}).  
Thus, three components of the total momentum are the integrals of 
motion, and the total momentum of the system of two particles is preserved.
	
By virtue of isotropy of the space, the Hamiltonian of a closed 
system must be invariant under a rotation of the whole system by an 
arbitrary angle around any axis.  It suffices to require the 
fulfillment of this condition for any infinitesimal rotation whose
vector $\delta \bm{\varphi}$ 
has the modulus equal to the rotation angle $\delta \varphi$
and is directed 
along the rotation axis.  The operator of the infinitesimal rotation 
transforming the function  $\Psi ({\bf r}_1 ,{\bf r}_2 )$ into 
$\Psi ({\bf r}_1 + [\delta \bm{\varphi} \times {\bf r}_1],\, {\bf r}_2
+ [\delta \bm{\varphi} \times {\bf r}_2 ])$ is connected with the operator of 
the total angular momentum of the system
\begin{equation}
1 + \delta \bm{\varphi}\cdot \sum\limits_{i = 1}^2 \left[ {\bf r}_i \times
\bm{\nabla}_i \right] = 1 + \frac{i}{\hbar }\delta \bm{\varphi}\cdot {\bf \hat
L}  \label{E63}
\end{equation}
and commutes with Hamiltonian~(\ref{E60}) of the system.  
Thus, the total angular momentum ${\bf \hat L} = - i\hbar 
\sum\limits_{i = 1}^2 \left[ {\bf r}_i
\times \bm{\nabla}_i \right] = - i\hbar \left[ {\bf r} \times \bm{\nabla}
_{\bf r} \right] - i\hbar \left[ {\bf R} \times \bm{\nabla}_{\bf
R}\right]$ of the system of two particles is preserved.
	
It is important to note the following fact.  
By writing formally the operators ${\bf \hat l}_1= \left[ 
{\bf \hat r}_1 \times {\bf \hat p}_1
\right]$ and ${\bf \hat l}_2 = \left[ {\bf \hat r}_2 
\times {\bf \hat p}_2\right]$ for 
each particle, we can easily prove that they and their 
sum are not angular momenta, because they do not satisfy 
the standard commutation relations intrinsic to the 
angular momentum:
\begin{equation}
\left[ {\hat L_x,\hat L_y } \right] = i\hbar \hat L_z,\;
\left[ {\hat L_y,\hat L_z } \right] = i\hbar \hat L_x,\;
\left[ {\hat L_z,\hat L_x } \right] = i\hbar \hat L_y\; .
\label{E64}
\end{equation}
However, from the operators ${\bf \hat r}_i$ and ${\bf \hat p}_j$,  we can 
construct an operator which will possess the 
above-mentioned properties of the total angular momentum:
\begin{equation}
{\bf \hat L} = \sum\limits_{i,j} {C_{ij} } \left[ {\bf \hat r}_i
\times {\bf \hat p}_j  \right] = - i\hbar \sum\limits_{i = 1}^2
\left[ {\bf r}_i \times \bm{\nabla}_i \right]\; .  \label{E65}
\end{equation}
For a system of two particles, the coefficients $C_{ij}$ read
\begin{eqnarray}
&&C_{11} = (1 - \varepsilon _{21} )\beta^{-1}\; ,  \label{E66}\\
&&C_{22} = (1 - \varepsilon _{12} )\beta^{-1}\; ,  \label{E67}\\
&&C_{12} = - \varepsilon _{21}\beta^{-1}\; , \label{E68}\\
&&C_{21} = - \varepsilon _{12}\beta^{-1}\; . \label{E69}
\end{eqnarray}

We should like to emphasize that the noncommutativity 
parameters of the operators of coordinates and momenta 
of different particles, $\varepsilon _{12}$ and $\varepsilon _{21}$, 
depend on the mean 
value of the modulus of the interaction force between 
two particles in the ground state (i.e., on the distance 
between the particles).  On the motion of two reference 
systems relatively each other with constant velocity ${\bf v}$,  
the operators ${\bf \hat r}_i$ and ${\bf \hat p}_i$ are 
transformed, respectively, into ${\bf \hat r}_i - {\bf v}t$ and 
${\bf \hat p}_i - m_i {\bf v}$.  
Such a Galilei transformation of a system of particles is 
described by the operator~\cite{R5}
\begin{equation}
\hat G({\bf v},t) = \exp\left[ {i{\bf v}(M{\bf \hat R} - {\bf \hat
P}t)/\hbar } \right]\; ,  \label{E70}
\end{equation}
where $M$ and ${\bf \hat P}$ are the mass and the operator of 
the total momentum of the system of two particles 
and ${\bf \hat R}$ is vector~(\ref{E59}) of a position of the fictitious 
free particle.  It is easy to show that the condition 
for the equation of motion~(\ref{E55}) to be Galilei-invariant,
\begin{equation}
\hat G^\dag ({\bf v},t)\left[ i\hbar \frac{\partial }{\partial t}
- H \right]\hat G({\bf v},t) = \left[ i\hbar \frac{\partial
}{\partial t} - H \right]\; , \label{E71}
\end{equation}
is satisfied for Hamiltonian (\ref{E60}).

For the wave function of relative motion of the particles,
we get the following equation:
\begin{equation}
\left[ - \frac{\hbar ^2 \beta^2}{2\mu }\Delta _{\bf r}  + V\left( \left| {\bf r} \right|
\right) \right]\phi \left( {\bf r} \right) = E\phi \left(
{\bf r} \right)\; ,  \label{E72}
\end{equation}
where
\begin{equation}
\beta = \frac{1 - \Omega
^2 \xi^2 \mu ^2 M^{-2} }  
{1 + \Omega ^2 \xi ^2
\mu ^2 M^{-2} + \Omega \xi \left( 1 - 2\mu / M
\right) }\; ,\label{E73}
\end{equation}
\begin{equation}
\xi = \frac{2\hbar }{\mu ^2 c^3 }\left\langle \phi _0
\right.\left| \left| {\bf F}_{12}  \right| \right|\left. \phi
_0  \right\rangle /\left\langle \phi _0 \left |\phi _0
\right. \right\rangle \; . \label{E74}
\end{equation}
The total energy of the system is $E_0 = E + E_{\bf R}$, where 
$E_{\bf R}$ is the energy 
of free motion of the first fictitious particle.
	
Similarly to the Schr\"{o}dinger nonrelativistic theory, the wave function  
$\phi ({\bf r})$ should be continuous together with its partial derivatives 
of the first order in the entire space.  In addition, the wave 
function $\phi ({\bf r})$ should be a bounded and one-valued function 
of its arguments.
	
Similarly to the Schr\"{o}dinger theory for particles interacting 
through a centrally symmetric potential which depends only 
on the distance between particles, the wave function $\phi ({\bf r})$  
can be represented as
\begin{equation}
\phi ({\bf r}) = \frac{1}{r}\chi _l (r)Y_{lm} \left( \frac{\bf
r}{r} \right)\; ,  \label{E75}
\end{equation}
where $Y_{lm} \left( {\bf n} \right)$ are the orthonormalized spherical functions.  
Then the function $\chi _l (r)$ satisfies the equation
\begin{equation}
\left[ - \frac{\hbar ^2 \beta^2 }
{2\mu }\left( \frac{d^2 }{dr^2 } - \frac{l(l + 1)}{r^2 }
\right) + V\left( r \right) \right]\chi_l (r) = E\chi_l (r)
\label{E76}
\end{equation}
which has solutions for a system of two interacting particles 
at definite values of the energy $E$.

The constant $\Omega$ can be determined by considering the ground 
state of the discrete spectrum of a hydrogenlike atom.  
Let two particles with masses $m_1$ (atomic nucleus) and $m_2$  
(electron) be bound by the Coulomb potential $V(r) = - Ze^2 /r$, where $Z$   
is the atomic nucleus charge. The equation for bound 
states can be written as
\begin{equation}
\left[ - \frac{\hbar ^2 \beta ^2} {2\mu }\left(
\frac{d^2 }{dr^2 } - \frac{l(l + 1)}{r^2} \right) - \frac{ Ze^2
}{r} \right]\chi _{nl} (r) = E_{nl} \chi _{nl} (r)\; . \label{E77}
\end{equation}
Here, we took into account that $m_1 \gg m_2 $ and then $\beta =(1 + \Omega 
\xi _0 )^{-1}$ and wrote the index 0 
to the parameter $\xi$ by emphasizing the fact that it 
is defined by the ground quantum state.
Equation~(\ref{E77}) is the equation for the radial functions 
of a hydrogenlike atom by the Schr\"{o}dinger theory whose 
normed solutions are well known for bound states 
(see, e.g.,~\cite{R7}):
\begin{eqnarray}
\chi _{nl} (r) = N_{nl}\, r^{l + 1} F\left( - n + l + 1,2l +
2,\frac{2Z r}{na_0 \beta ^2} \right)\nonumber \\
\times \exp\left(-
\frac{Z r}{na_0 \beta ^2} \right)\; ,  \label{E78}
\end{eqnarray}
where
\begin{equation}
N_{nl} = \frac{1}{(2l + 1)!}\left[\frac{(n + l)!}{2n(n - l - 1)!}
\right]^{1/2}\left( \frac{2Z}{na_0 \beta^2 }
\right)^{l+3/2}\; . \label{E79}
\end{equation}
Here, $a_0 = \hbar ^2 /\mu e^2$  is the Bohr radius and $F$ is a 
confluent hypergeometric function.  
The eigenvalues of the energy of bound states are
\begin{equation}
E_{nl} = - \frac{\mu c^2 }{2} \frac{(\alpha Z)^2 }{n^2
}\left( 1 + \Omega \xi _0  \right)^2\; .  \label{E80}
\end{equation}
Here, $l = 0,1,\dots,n - 1$ and $n = 1,2,\dots,\infty$.
	
By substituting $\chi _{10} (r)$ into Eq.~(\ref{E74}), we obtain the nonlinear 
equation for the determination of $\xi _0$:
\begin{equation}
\frac{\eta _0 }{(1 + \eta _0 )^4 } = 4\Omega (\alpha Z)^3;\;\;\;\;
\eta _0 = \Omega \xi _0\; .  \label{E81}
\end{equation}
\begin{figure}[tbp]
\centering
\includegraphics[width=4in]{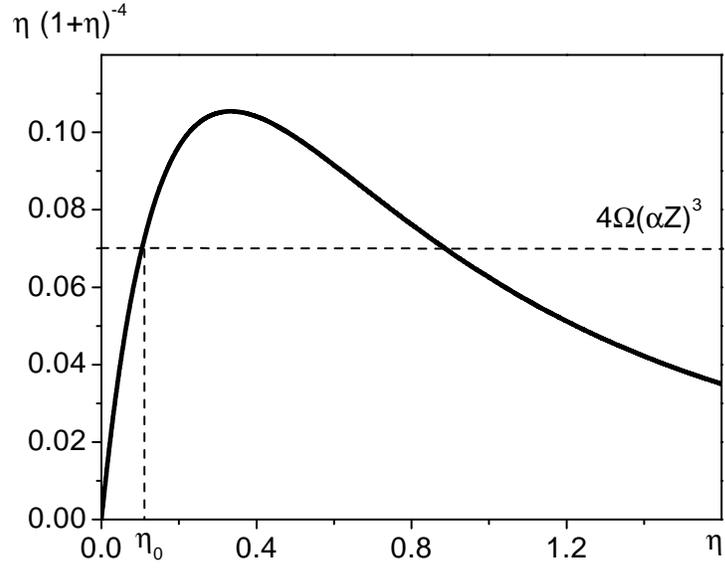}
\caption{Graphical solution of Eq.~(\ref{E81}) for $\protect\eta _0$  
at various values of the parameter $4\Omega (\protect\alpha Z)^3$ 
for the ground state of a hydrogenlike atom.}
\label{fig1}
\end{figure}
The nonlinear equation~(\ref{E81}) for $\eta _0$ has solutions under 
the condition $4\Omega (\alpha Z)^3 \le 27/256$,  
which is shown in Fig.~\ref{fig1}. From two solutions, 
suitable is a solution which is positioned nearer to zero.  
The second solution should be omitted since it corresponds 
to the case where the binding energy increases with decrease 
in the parameter $\alpha Z$. For $4\Omega (\alpha Z)^3 > 27/256$,
Eq.~(\ref{E81}) has no solutions, 
which means the impossibility for a given bound state to exist.  
The critical value of the interaction constant $Z = Z_C$ is reached 
at $\eta _0 = 1/3$. In this case, the mean distance between particles attains 
the minimum value
\begin{equation}
\left\langle \left| {\bf r}_2 - {\bf r}_1  \right|
\right\rangle  = \frac{3}{2}\frac{\hbar }{\mu c}\frac{1}{\alpha
Z_C } \frac{1}{(1 + \eta _0 )^2}\; . \label{E82}
\end{equation}
\begin{figure}[tbp]
\centering
\includegraphics[width=4in]{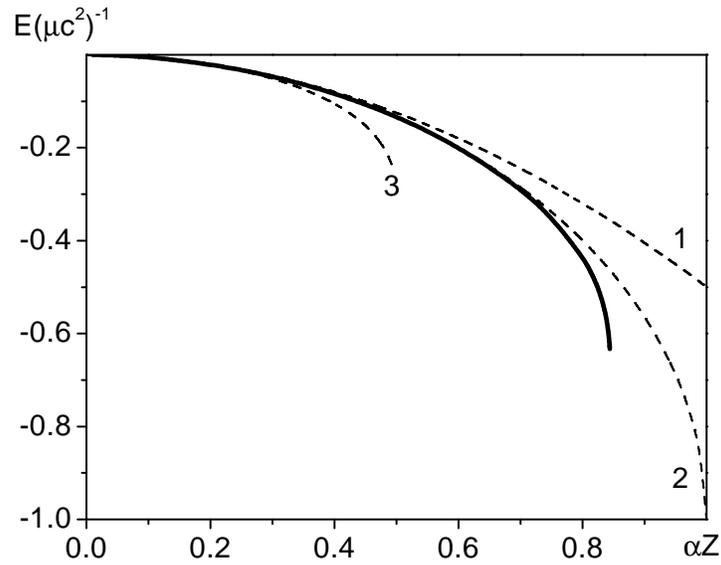}
\caption{Binding energy of the ground state of a 
hydrogenlike atom vs the parameter $\protect\alpha Z $.  
For the sake of comparison, we present the corresponding 
dependences according to the equations of Schr\"{o}dinger~(1), 
Dirac~(2), and Klein-Gordon~(3).}
\label{fig2}
\end{figure}
By requiring that this least value be equal to $\hbar /\mu c$,  
we determined the parameter $\Omega$ whose value is $32/729$.  
In this case, the critical value of $Z_C$ equals $115.6\;\;  
\left( \alpha Z_C = 27/32 \right)$.
The binding energy for the ground state of a hydrogenlike atom, 
which is calculated by using the parameter $\Omega$ determined in such 
a way, is displayed in Fig.~\ref{fig2}, where the analogous dependences of 
the binding energies within the Schr\"{o}dinger, Dirac, and Klein-Gordon 
theories are also presented.  As seen, the energy levels of the 
ground state are positioned below the Schr\"{o}dinger levels and above 
those calculated by the Dirac theory in a rather wide interval of 
values of the interaction constant ($0 < \alpha Z < 0.685$). 
Excited states of a 
hydrogenlike atom are positioned below relevant Schr\"{o}dinger levels.
	
The noncommutativity parameters of the operators of coordinates 
and momenta of different particles $\varepsilon _{12}$ and $\varepsilon _{21}$
for a hydrogenlike atom can be estimated as follows:
\begin{equation}
\varepsilon _{12} = \Omega \xi _0 \frac{m_2^2 }{M^2} = \eta _0
\frac{m_2^2}{M^2}\; ,  \label{E83}
\end{equation}
\begin{equation}
\varepsilon _{21} = \frac{\Omega \xi _0 }{1 + \Omega \xi _0 } =
\frac{ \eta _0 }{1 + \eta _0 }\; .  \label{E84}
\end{equation}
\begin{figure}[tbp]
\centering
\includegraphics[width=4in]{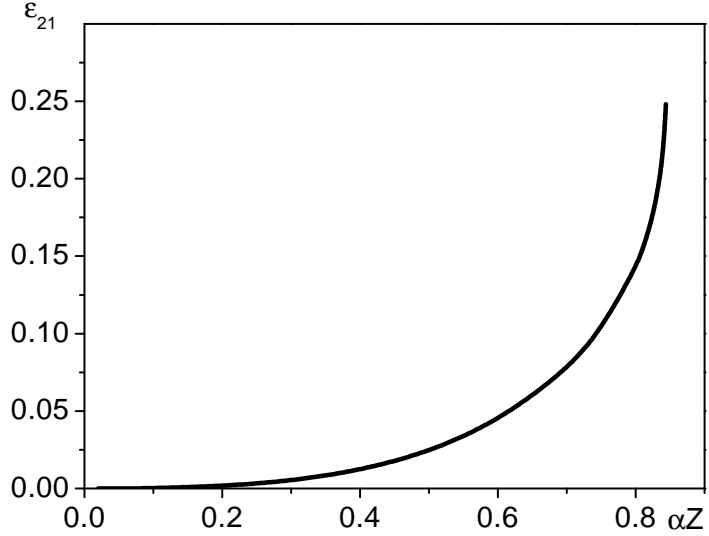}
\caption{Dependence of the noncommutativity parameter $\protect\varepsilon _{21}$ 
on the interaction constant for a hydrogenlike atom.}
\label{fig3}
\end{figure}The last dependence is shown in Fig.~\ref{fig3}. For a hydrogen atom, 
the noncommutativity parameters are $\varepsilon _{12} = 2.0 \times 10^{ - 14}$ 
and  $\varepsilon _{21} = 6.8 \times 10^{ - 8}$.

Of significant interest is the comparison of the obtained results with the 
available experimental data and the consequences of the Schr\"{o}dinger nonrelativistic 
theory because the high-precision experimental measurements 
of the energy levels of a hydrogen atom ~\cite{RR3}, light hydrogenlike atoms 
(see the review ~\cite{RR5}), and heavy ions $^{92}U$ with one~\cite{RR4} or several 
electrons~\cite{RR1,RR2} have been recently performed.

Table~\ref{tab1} presents the ground state energies of some hydrogenlike atoms together 
with experimental data and the results following from the Schr\"{o}dinger equation. 
The experimental data for $Z=6 - 42$ and  
$Z=92$ are taken, respectively, from~\cite{RR6} and~\cite{RR4}.  
The last two columns, in which the differences 
of the theoretical energy levels by Schr\"{o}dinger and by Eq.~(\ref{E80}) with 
experimental data are given, 
demonstrate the advantage of the proposed nonrelativistic quantum-mechanical 
method of description 
of hydrogenlike atoms at great interaction constants $\alpha Z$.  
Indeed, the difference between the theoretical 
and experimental values of the ground state energy of a hydrogenlike atom for 
middle values of $Z$  
is approximately twice less than that by Schr\"{o}dinger.  
The very good agreement with the experimental 
value of the ground state energy is obtained for hydrogenlike uranium, which 
corresponds to the 
region of intersection  ($Z = 94$) of the theoretical curve of the ground state 
energy versus  $\alpha Z$  
and the relevant curve (Fig.~\ref{fig2}) for the Dirac equation.  In the region of the 
critical value of the 
interaction constant  ($Z = 115$), the significant role is played by relativistic 
effects.  Therefore, in this 
case, one should expect a worse agreement with experimental data.  In addition, 
the consideration 
of relativistic effects can change the critical value of the interaction constant 
in the direction of its growth.  
Analogous conclusions can be drawn from Table~\ref{tab2} which gives the theoretical, 
experimental, and 
Schr\"{o}dinger-equation-based values of the gap between levels 1s and 2s.

\begin{table}
\caption{\label{tab1}
Binding energy of the ground state $E$ (see Eq.~(\ref{E80})) for certain hydrogenlike 
atoms as compared to the experimental data  $E_{\rm{EXP}}$ and those by 
the Schr\"{o}dinger equation $E_S$.  All values are given in $\mathop{\rm eV}$.}
\begin{ruledtabular}
\begin{tabular}{l D{.}{.}{4} D{.}{.}{4} D{.}{.}{4} D{.}{.}{4}}
\multicolumn{1}{c}{$Z$}&\multicolumn{1}{c}{$E$}&
\multicolumn{1}{c}{$E_{\rm{EXP}}$}&\multicolumn{1}{c}{$E_S-E_{\rm{EXP}}$}&
\multicolumn{1}{c}{$E-E_{\rm{EXP}}$}\\
\hline
6  & -489.8193 & -489.9933 & 0.1884 & 0.1740 \\
12 & -1959.682 & -1962.665 & 3.445  & 2.983  \\
18 & -4411.759 & -4426.224 & 17.980 & 14.465 \\
24 & -7851.73  & -7894.80  & 57.92  & 43.07  \\
30 & -12290.62 & -12388.93 & 143.81 & 98.31  \\
36 & -17746.88 & -17936.21 & 303.23 & 189.33 \\
42 & -24248.77 & -24572.23 & 571.79 & 323.46 \\
92 & -131726.  & -131812.  & 16653. & 86.    \\
\end{tabular}
\end{ruledtabular}
\end{table}

\begin{table}
\caption{\label{tab2}
Gaps between levels 1s and 2s for certain hydrogenlike atoms calculated in this 
work $\Delta=E(2s)-E(1s)$ as compared to the experimental data $\Delta_{\rm{EXP}}$ 
and those by the Schr\"{o}dinger equation $\Delta_S$.  
All values are given in $\mathop{\rm eV}$.}
\begin{ruledtabular}
\begin{tabular}{l D{.}{.}{4} D{.}{.}{4} D{.}{.}{4} D{.}{.}{4}}
\multicolumn{1}{c}{$Z$}&\multicolumn{1}{c}{$\Delta$}&
\multicolumn{1}{c}{$\Delta_{\rm{EXP}}$}&\multicolumn{1}{c}
{$\Delta_{\rm{EXP}}-\Delta_S$}&
\multicolumn{1}{c}{$\Delta_{\rm{EXP}}-\Delta$}\\
\hline
6  & 367.3645 & 367.4774 & 0.1237 & 0.1129 \\
12 & 1469.761 & 1471.729 & 2.314  & 1.968  \\
18 & 3308.819 & 3318.338 & 12.155 & 9.519  \\
24 & 5888.794 & 5916.929 & 39.270 & 28.135 \\
30 & 9217.96  & 9281.538 & 97.696 & 63.578 \\
36 & 13310.16 & 13431.01 & 206.28 & 120.85 \\
42 & 18186.58 & 18389.67 & 389.34 & 203.09 \\
\end{tabular}
\end{ruledtabular}
\end{table}

\begin{figure}[tbp]
\centering
\includegraphics[width=4in]{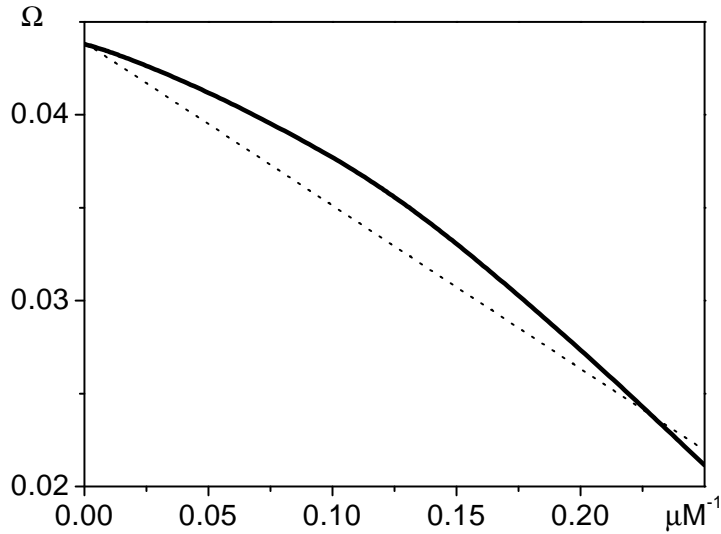}
\caption{Parameter $\Omega $ vs $\protect\mu /M$ in the two-body problem.}
\label{fig4}
\end{figure}

The constant $\Omega = 32/729$ is derived under the 
condition $\mu /M \ll 1$.  
For another relation between the masses of interacting 
particles, it is necessary to use the complete expression 
for $\beta$~[ Eq.~(\ref{E73})] 
to derive the constant $\Omega$ from 
the condition that the minimum mean distance between 
particles reaches $\Delta _{12}$. This dependence is shown in 
Fig.~\ref{fig4}.  
The good approximation of the dependence of $\Omega$ on $\mu /M$ is 
attained by the expression $\Omega = 32(1 - 2\mu M^{-1} )/729$ 
shown in Fig.~\ref{fig4} by the 
dotted line.  This approximation is convenient for a 
quantum system composed of several particles with different masses.  
Of great interest is the situation with two identical 
particles.  In this case, $\mu /M = 0.25$, $\beta = \left(1
- \Omega \xi _0 /4 \right)/\left(1 + \Omega \xi _0 /4 
\right)$, and $\Omega = 0.0211547$.
	
For other interaction potentials between particles, 
we may take the parameter $\Omega$ which was derived for a 
hydrogenlike atom.  We note that even the potentials 
with a singularity at zero (those of the Yukawa or Hulth\'{e}n type) 
lead to the mean distance between particles which is at 
least $\Delta _{12}$ given by Eq.~(\ref{E2}).
	
Below, we give the values of the Poisson quantum brackets 
proposed by Dirac~\cite{R8}:
\begin{eqnarray}
&&\left\{ \hat x_1\, ,\hat p_{1x}  \right\} = 1 - \varepsilon
_{12}\; , \label{E85}\\
&&\left\{ \hat x_2\, ,\hat p_{2x} \right\} = 1 - \varepsilon _{21}\;
, \label{E86}\\
&&\left\{ \hat x_1\, ,\hat p_{2x} \right\} = \varepsilon _{12}\; ,
\label{E87}\\
&&\left\{ \hat x_2\, ,\hat p_{1x} \right\} = \varepsilon _{21}\; ,
\label{E88}\\
&&\left\{ \hat x_1\, ,\hat x_2  \right\} = 0\; ,  \label{E89}\\
&&\left\{ \hat p_{1x}\, ,\hat p_{2x}  \right\} = 0.  \label{E90}
\end{eqnarray}
For $\varepsilon _{12} ,\varepsilon _{21} \to 0$,  these brackets 
are transformed into the classical 
Poisson ones, i.e., we have the full analogy between classical 
mechanics and quantum one in this case.  As seen in Fig.~\ref{fig3},
$\varepsilon _{21}$
differs considerably from zero in the systems whose sizes are of 
order of the Compton wavelengths of the particles composing a system.  
In this case, the analogy with classical mechanics is absent.

\section{Nonrelativistic system of $N$ interacting particles}

The above results can be easily generalized for a system 
consisting of $N$ particles which are bound with one another 
by two-particle forces.

Let the operators of coordinates and momenta of $N$ particles 
be ${\bf \hat r}_1$, ${\bf \hat r}_2$, $\dots$ , ${\bf \hat r}_N$, 
${\bf \hat p}_1$, ${\bf \hat p}_2$, $\dots$ , ${\bf \hat p}_N$.  
We define the operator of the total momentum of the system as
\begin{equation}
{\bf \hat P} = \sum\limits_{k = 1}^N {\bf \hat p}_k\; .
\label{E91}
\end{equation}
Analogously to the two-particle problem, we require that the 
commutator of the coordinate operator of any particle with the 
operator of the total momentum of the system be equal to $i\hbar$:
\begin{equation}
\left[ {\bf \hat r}_i\, ,{\bf \hat P} \right] = i\hbar\; ,\;\;\; i
= 1,2,\dots,N\; . \label{E92}
\end{equation}
Then if
\begin{equation}
\left[ {\bf \hat r}_i\, ,{\bf \hat p}_k  \right] = i\hbar
\varepsilon _{ik}\; ,\;\;\; i \ne k \; , \label{E93}
\end{equation}
we get
\begin{equation}
\left[ {\bf \hat r}_i\, ,{\bf \hat p}_i  \right] = i\hbar \left[ 1
- \sum\limits_{k = 1}^N \varepsilon _{ik} \right]\; ,\;\;\; i =
1,2,\dots,N\; . \label{E94}
\end{equation}
Here, the noncommutativity parameter of the operators of 
coordinates and momenta of different particles $\varepsilon _{ik}$ equals zero 
for $i = k$ ($\varepsilon _{ii} = 0$), which was made for the sake 
of convenience to write 
the further formulas. In addition,
\begin{equation}
\left[{\bf \hat r}_i\, ,{\bf \hat r}_k  \right] = 0\; ,\quad
\left[{\bf \hat p}_i\, ,{\bf \hat p}_k  \right] = 0\; .\label{E95}
\end{equation}

One of the possible representations of the operators of 
coordinates and momenta of particles can be written as
\begin{equation}
{\bf \hat r}_i = {\bf r}_i\; ,  \label{E96}
\end{equation}
\begin{equation}
{\bf \hat p}_i = - i\hbar \left[1 - \sum\limits_{s = 1}^N
\varepsilon _{is} \right]\bm{\nabla}_i - i\hbar \sum\limits_ {k = 1}^N
\varepsilon _{ki} \bm{\nabla}_k\; ,
\label{E97}
\end{equation}
where $i = 1,2,\dots,N$. 	
Here, we use the coordinates of particles as independent 
variables, because the relevant operators are commutative.
	
In this case, the equation for the nonrelativistic $N$-particle 
problem takes the form
\begin{eqnarray}
\left\{- \frac{\hbar ^2}{2}\sum\limits_{i = 1}^N \left[\frac{A_i}
{m_i}\Delta _i + \sum\limits_{k > i}^N 2B_{ik} \left( \bm{\nabla}_i
\cdot \bm{\nabla}_k \right) \right]\right. \nonumber \\
+\left. \sum\limits_{j > i}^N V \left(
\left| {\bf r}_i - {\bf r}_j \right| \right) \right\}\Psi =
E\Psi\; , \label{E98}
\end{eqnarray}
where
\begin{equation}
A_i = \left(1 - \sum\limits_{s = 1}^N \varepsilon _{is} \right)^2
+ \sum\limits_{s = 1}^N \frac{m_i}{m_s}\varepsilon _{is}^2\; ,
\label{E99}
\end{equation}
\begin{equation}
B_{ik} = \frac{\varepsilon _{ki}}{m_i }+\frac{\varepsilon _{ik}}{m_k }
+ \sum\limits_{s=1}^N\left(\frac{\varepsilon _{ks}\varepsilon _{is}}{m_s } 
-\frac{\varepsilon _{ki}\varepsilon _{is}}{m_i }
-\frac{\varepsilon _{ik}\varepsilon _{ks}}{m_k }\right). 
\label{E100}
\end{equation}

By using the transformation
\begin{equation}
{\bf R}_1 = \frac{m_1 {\bf r}_1 }{m_1} - {\bf r}_2\; ,
\label{E101}
\end{equation}
\begin{equation}
{\bf R}_2 = \frac{m_1 {\bf r}_1 + m_2 {\bf r}_2 }{m_1 + m_2 } -
{\bf r}_3\; , \label{E102}
\end{equation}
\begin{equation}
{\bf R}_{N - 1} = \frac{m_1 {\bf r}_1 + \dots + m_{N - 1} {\bf r}_{N
- 1} }{ m_1 + \dots + m_{N - 1} } - {\bf r}_N\; ,  \label{E103}
\end{equation}
\begin{eqnarray}
{\bf R}_N & = & \frac{m_1 {\bf r}_1 + \dots + m_N {\bf r}_N }{m_1 +
\dots + m_N }-(a_1 + \dots + a_{N - 1} ) {\bf r}_N\nonumber \\
&&+ a_1 {\bf r}_1 + \dots + a_{N - 1} {\bf r}_{N - 1}\; ,\label{E104}
\end{eqnarray}
we can separate the free motion of some fictitious particle 
whose mass is equal to the mass of all the system, 
$M = \sum\limits_{k = 1}^N {m_k }$.  
The first $N - 1$ equations correspond to the well-known Jacobi 
transformation of coordinates.  Besides the coordinate of 
the center of masses of the system, the last equation includes 
the additional terms with $N - 1$ unknown parameters whose values 
can be determined from the condition that the coefficients of 
mixed derivatives in the operator of kinetic energy, 
$\left( \bm{\nabla}_{{\bf R}_1} \cdot \bm{\nabla}_{{\bf R}_{N}} \right)$,
$\left( \bm{\nabla}_{{\bf R}_2} \cdot \bm{\nabla}_{{\bf R}_{N}} \right)$,
$\dots$ , $\left( \bm{\nabla}_{{\bf R}_{N-1}} \cdot \bm{\nabla}_{{\bf R}_{N}}
\right)$,  
are equal to zero, i.e., $N - 1$ equations allow one to determine $N - 1$  
unknown parameters.  In the general case, the expressions for 
the parameters $a_1$, $a_2$, $\dots$, $a_{N - 1}$ are cumbersome.  
Therefore, in addition 
to formulas~(\ref{E58}) and (\ref{E59}) of the two-body problem, we present 
only the values of parameters for the three-body problem:
\begin{eqnarray}  \label{E105}
a_1 = \frac{m_1}{M d}\left[ \varepsilon _{13} (1 - \varepsilon
_{21} - \varepsilon _{23}- \varepsilon _{32} )\right] \nonumber \\
+ \frac{m_1}{M d}\left[ \varepsilon
_{12} (1 - \varepsilon _{23} - \varepsilon _{31} - \varepsilon
_{32} ) \right] \nonumber \\ 
+ \frac{m_2}{M d}\left[ \varepsilon _{21} ( - 1
+ \varepsilon _{31} + \varepsilon _{32} ) + \varepsilon _{23}
\varepsilon _{31} \right] \nonumber \\ 
+ \frac{m_3}{M d}\left[\varepsilon
_{31} ( - 1 + \varepsilon _{21} + \varepsilon _{23} ) +
\varepsilon _{21} \varepsilon _{32}  \right]\; ,
\end{eqnarray}

\begin{eqnarray}  \label{E106}
a_2 = \frac{m_2}{M d}\left[ \varepsilon _{21} (1 - \varepsilon
_{31} - \varepsilon _{32} - \varepsilon _{13} )\right]\nonumber\\ 
+ \frac{m_2}{M d}\left[\varepsilon_{23} (1 - \varepsilon _{31} 
- \varepsilon _{12} - \varepsilon_{13} ) \right] \nonumber\\ 
+ \frac{m_1}{M d}\left[\varepsilon _{13}
\varepsilon _{32} + \varepsilon _{12} ( - 1 + \varepsilon _{31} +
\varepsilon _{32} ) \right] \nonumber\\ 
+ \frac{m_3}{M d}\left[ \varepsilon
_{32} ( - 1 + \varepsilon _{13} + \varepsilon _{12} ) +
\varepsilon _{12} \varepsilon _{31}  \right]\; .
\end{eqnarray}
Here,  
$d = ( - 1 + \varepsilon _{21} + \varepsilon _{23} )( - 1 + \varepsilon
_{31} + \varepsilon _{13} ) + \varepsilon _{32} ( - 1 + \varepsilon _{21} +
\varepsilon _{13} ) + \varepsilon _{12} ( - 1 + \varepsilon _{23} +
\varepsilon _{31} + \varepsilon _{32} )$.

The system of equations~(\ref{E98})-(\ref{E100}) takes the especially 
simple form in the important case of identical particles  ($m_i = m$,
$\varepsilon _{ij} = \varepsilon$, $i,j = 1,\dots,N$, $i \ne j$)  
after the introduction of normed Jacobi coordinates
\begin{equation}
{\bf q}_k = \sqrt {\frac{k}{k + 1}} \left(\frac{1}{k}
\sum\limits_{s=1}^k {\bf r}_s - {\bf r}_{k + 1} \right)\; , \; 1 \le
k \le (N - 1)\; , \label{E107}
\end{equation}
\begin{equation}
{\bf q}_N = \frac{1}{\sqrt N }\sum\limits_{s = 1}^N {\bf r}_s\; .
\label{E108}
\end{equation}
In this case, after the separation of the free motion of a fictitious 
particle whose mass is equal to the mass of the whole system, we get
\begin{eqnarray}
\left\{- \frac{\hbar ^2 (1 - N\varepsilon )^2 }{2m}(\Delta_{{\bf
q} _1}+ \dots + \Delta_{{\bf q}_{N - 1}} ) \right. \nonumber \\
+ \left. V({\bf q}_1 ,\dots,{\bf q}_{N - 1} ) \right\}\phi  =
E\phi\; ,  \label{E109}
\end{eqnarray}
\begin{equation}
\varepsilon = \frac{\kappa }{1+\kappa }\;,\;\;\; 
\kappa =\frac{2\hbar \Omega
\left(0.25\right){\left\langle \phi _0
\left|\left|{\bf F}_{12} \right|\right|\phi _0 \right
\rangle}}{m^2 c^3 \left\langle \phi _0
\left|\right.\phi _0 \right\rangle} \; . \label{E110}
\end{equation}
Here, $V({\bf q}_1 ,{\bf q}_2 ,\dots,{\bf q}_{N - 1} )$ 
is the potential energy of the  $N$-particle system.

\section{Conclusion}

The Schr\"{o}dinger equation for a system of interacting particles 
is not strictly nonrelativistic since it is based on the implicit 
assumption that the interaction propagation velocity is finite.  
The last means that, if the commutator of the operators of the 
coordinate and the relevant momentum of a free particle is
\begin{equation}
\left[ {\hat x\, ,\;\hat p_x } \right] = i\hbar\; ,  \label{E111}
\end{equation}
then this commutator takes the same value $i\hbar$ for a system 
of bound particles.  However, in a nonrelativistic quantum system, 
the total momentum transferred is distributed upon the measurement 
of the coordinate of a particle over all the particles rather than 
it is transferred to the measured particle.  Therefore, in a system 
of interacting particles, this commutator must have the form
\begin{equation}
\left[ {\hat x\, , \, \hat p_x } \right] = i\hbar \delta\; ,
\label{E112}
\end{equation}
where $0 < \delta < 1$.

The refusal of the implicit assumption about the finiteness of 
the interaction propagation velocity leads to the noncommutativity 
of the operators of coordinates and momenta of different particles. 
However, the operators of coordinates of all the particles as well 
as the operators of momenta of all the particles are commutative, 
which allows one to use these collections as independent variables.
	
The properties of solutions of the proposed equation differ 
considerably from those of the Schr\"{o}dinger solutions for the 
systems in which the Compton wavelengths of particles are 
comparable with the system size.  That is, the consideration of 
the noncommutativity of the operators of coordinates and momenta 
of different particles is important for the quantum mechanics of 
atoms with large charge of a nucleus ($\alpha Z\approx 1$)  
and for the phenomena of 
nuclear physics where the size of a system is of order of the 
Compton wavelengths of particles composing the system.
	
\begin{acknowledgments}
The author thanks sincerely V.~V.~Kukhtin and 
A.~I.~Steshenko for the useful discussions.
\end{acknowledgments}
\bibliography{kuzmenko}

\end{document}